\documentclass[12pt]{article}
\input epsf.tex
\def\DESepsf(#1 width #2){\epsfxsize=#2 \epsfbox{#1}}
\usepackage{epsfig}
\usepackage{color}
\usepackage{array}
\makeatletter
\def\alt{\mathrel{\mathpalette\gl@align<}}
\def\agt{\mathrel{\mathpalette\gl@align>}}
\def\gl@align#1#2{\lower.6ex\vbox{\baselineskip\z@skip\lineskip\z@
\ialign{$\m@th#1\hfil##\hfil$\crcr#2\crcr\sim\crcr}}}
\makeatother

\begin{document}
\begin{flushright}
{\tt hep-ph/0406262}\\
June, 2004 \\
UMD-PP-04-045\\
\end{flushright}
\vspace*{2cm}
\begin{center}
{\baselineskip 25pt \large{\bf Neutrino Masses and Mixings in a Predictive
SO(10) Model \\with CKM CP Violation} \\
}

\vspace{1cm}

{\large
Bhaskar Dutta{$^*$},
Yukihiro Mimura{$^*$}
and R.N. Mohapatra{$^\dagger$}

\vspace{.5cm}

{\it 
$^*$Department of Physics, University of Regina, \\
Regina, Saskatchewan S4S 0A2, Canada \\
$^\dagger$Department of Physics, University of Maryland, \\
College Park, MD 20742, USA\\
}}
\vspace{.5cm}

\vspace{1.5cm}
{\bf Abstract}
\end{center}

It has recently been shown that a minimal SO(10) model with a single {\bf
10} and a single {\bf 126} Higgs field breaking B-L symmetry
 predicts large solar and atmospheric mixings
in agreement with observations if it is assumed that the neutrino mass
obeys the type II seesaw formula. No additional symmetries need to be
assumed for this purpose.  Understanding CP violation
in the renormalizable version of the model however requires a significant 
non-CKM source. In this paper we show that if we extend the
model by the
inclusion of a heavy {\bf 120} dimensional Higgs field, then it can
accommodate CKM CP violation while remaining predictive in the neutrino
sector. Among the predictions are: (i) solar mixing angle in the observed
range; (ii) $\theta_{13}$ in the range of $0.1$
to $0.26$; (iii) the Dirac phase close to maximal for the central value of
the solar mixing angle.

\thispagestyle{empty}

\bigskip

\newpage

\section{Introduction}
The simplest grand unified model for understanding small neutrino masses
appears to be the SO(10) model\cite{so10} for the following reasons: (i)
it automatically brings in the right handed neutrino, $N_R$,  needed to
implement
the seesaw\cite{seesaw} mechanism since it fits in with other
standard model fermions in the {\bf 16} dimensional spinor representation
(ii) it contains the $SU(4)_c$ symmetry\cite{ps} which relates the quark
and lepton coupling parameters and in turn helps the predictivity of the
model in the neutrino sector by reducing the number
of parameters; (iii) it also contains the B-L symmetry\cite{ps,BL} needed
to keep the right handed neutrino masses below the Planck scale and
provides a group theoretic explanation of why neutrinos are necessarily
Majorana particles.

While all these make the SO(10) models appealing for neutrino mass studies,
detailed quantitative predictions generally involve too many parameters
limiting the predictive power unless extra symmetries (e.g. family
symmetries etc) are imposed on the theory. One
exception to this is the class of models that uses only one {\bf 10} and
one {\bf 126} Higgs multiplet to generate fermion
masses\cite{babu}. The original set of papers on this
model\cite{babu,others} used type I seesaw formula for neutrino mass is
given by
${\cal M}_\nu ~=~ -M_\nu^DM^{-1}_R(M_{\nu}^D){}^T$ where $M_\nu^D$ is the Dirac
mass of the neutrinos and $M_R$ is the mass matrix of the right handed
neutrinos. These predictions are now in contradiction with
experiments. 

It was subsequently pointed out in ref.\cite{goran} that if one uses type
II seesaw formula for the neutrino masses\cite{seesaw2} instead, the model
automatically predicts large atmospheric mixing angle due to the fact
that bottom quark and tau lepton masses converge towards each other
when extrapolated to the GUT scale. The question remained whether this
works for three generations and can lead to a realistic model for
neutrinos. It was shown in ref.\cite{goh} that the same $b-\tau$ mass
convergence  not only leads to a large solar mixing angle, but also
to a small and detectable value for $U_{e3}\equiv \sin\theta_{13}$. A detailed
numerical analysis was carried out that showed that the model is indeed in
agreement with present neutrino data and in particular the prediction of a
``large'' value for $\theta_{13}$ which makes this
model testable at the current as well as at the proposed long base line neutrino
experiments.

In the three generation neutrino discussion in ref.\cite{goh}, the Yukawa
couplings of fermions
were assumed to be real and all CP violating effects were assumed to
originate from the supersymmetry breaking sector. It is
however interesting to check if
one can accommodate the CKM
phase in the model by introducing phases
in the couplings. A detailed investigation of
the minimal
model where CP violation is introduced through complex Yukawa couplings 
(as in the standard model) showed\cite{goh1} that compatibility with
neutrino data
requires the CKM phase to be outside the first quadrant
whereas the standard model  CKM phase is in the first
quadrant\cite{buras}. This would
seem to imply that in order to understand observed CP violation in this
model, one must invoke a significant non-CKM source for CP violation (as
in the model with real Yukawa couplings) e.g. 
CP violation from the supersymmetry breaking sector. This could very well
be true. However, since all observed CP violating phenomena seem to be
explainable by the CKM model, it is important to
see whether one can explain both CKM CP violation and neutrino mixings by
a minimal modification of this SO(10)
model. There are also other issues such as SUSY CP problem that one needs
to address in the context of supersymmetry and it would be interesting to
see how these can be addressed in this model.

In this paper, we propose a very minimal way to incorporate CP violation
into the model, which not only leads to a predictive model in the neutrino
sector but also seems to have wider
implication beyond just explaining CKM CP violation. For instance, the
model presents a solution to the SUSY CP problem.

In order to attain our goal, we include a heavy {\bf
120} field with an extra $Z_2$ symmetry which we will call ``parity''  
symmetry imposed on
the model\footnote{Very different extensions of the model that use {\bf
120} but no symmetry have been discussed in
ref.\cite{osimo}.}. At energy scales below the mass of the {\bf 120}
field, the effect of this field is to appear as a higher dimensional
contribution to the Yukawa couplings. This effective theory has the
following properties. Despite the fact there are now three extra
parameters in the model, the theory still remains predictive in the
neutrino sector. Secondly, the mass
matrices for quarks and leptons are hermitian, which
therefore has the potential to solve other problems of supersymmetric
models such as the SUSY CP problem. In this paper we focus only on the
neutrino sector.

The main results of this paper are as follows: (i) using
type II seesaw formula we are able to
accommodate the CKM CP phase while keeping the model predictive
in the neutrino sector; for example, we predict the solar mixing angle in
the right range and $U_{e3}\geq 0.1$; (ii) the model has the potential to
solve the SUSY CP problem and
 (iii) it predicts the Dirac phase of PMNS matrix to be near maximal
for the central value of the solar mixing angle
tan$^2\theta_\odot\simeq 0.4$. 

The paper is organized as follows: in sec. 2, we introduce the model with
the inclusion of the {\bf 120} Higgs field and write down the
 fermion mass formulae in the general case;
in sec. 3, we impose parity symmetry on the model making it 
predictive in the neutrino sector; in sec. 4, we discuss the predictions
 for neutrino mixings and Dirac CP phase for neutrinos; in sec. 5, we
present our conclusions and discuss the outlook for the model.

\section{ SO(10) Model and CP violation}
We start by writing down the Yukawa interactions of our model, which are
responsible for the discussion of neutrino masses and mixings.
The Yukawa superpotential involves the  couplings of the
$\mathbf{16}$-dimensional
matter spinor $\psi_i$ with $\mathbf{10}$ ($H$), $\overline{\mathbf{126}}$
($\overline\Delta$), and $\mathbf{120}$($A$)  dimensional Higgs fields:
\begin{equation}
W_Y = \frac12 h_{ij} \psi_i \psi_j H  + \frac12 f_{ij} \psi_i \psi_j
\overline\Delta  + \frac12 h'_{ij} \psi_i \psi_j A .
\end{equation}
The Yukawa couplings, $h$ and $f$, are symmetric matrices, whereas $h'$ is
an anti-symmetric matrix due to SO(10) symmetry.
They are all complex matrices in general.

Once the SO(10) symmetry breaks down to the standard model symmetry,
we have four pairs of Higgs doublets arising from the $H$,  
$\overline\Delta$, and $A$ Higgs fields. There may also be other Higgs
doublets e.g. in {\bf 210} multiplet.
Under the $G_{422}= SU(4)_c\times SU(2)_L\times SU(2)_R$  decomposition we
have the following representations that contain the Higgs
doublets of up and down type: one pair arises from $H \supset
(\mathbf{1,2,2})$,
one pair comes from $\overline\Delta \supset (\mathbf{15,2,2})$,
and two pairs come form $A \supset (\mathbf{1,2,2})+(\mathbf{15,2,2})$.
We assume that one pair of their linear combinations, $H_u$ and $H_d$,
remains massless (mass is $\sim$ $O(v_{wk})$) and become the MSSM Higgs
doublets. As for other pairs, they all have GUT scale masses.
Using the light Higgs doublets, the MSSM Yukawa couplings below the GUT
scale and the right-handed Majorana neutrino 
mass terms can be written as
\begin{equation}
W_Y \supset
Y_{ij}^u Q_i U^c_j H_u + Y_{ij}^d Q_i D^c_j H_d + Y_{ij}^e L_i E^c_j H_d
+ Y_{ij}^\nu L_i N^c_j H_u
+ \frac12 f_{ij} L_i L_j \overline\Delta_L + \frac12 f_{ij} N^c_i N^c_j  
\overline\Delta_R^0,
\end{equation} 
where $Q,U^c,D^c,L,E^c,N^c$ are the quark and lepton superfields which
are all unified to the $\mathbf{16}$
spinor $\psi$ field.  $\overline\Delta_L$ is an $SU(2)_L$ triplet Higgs field
and $\overline\Delta_R^0$ is a neutral component of $SU(2)_R$
triplet, both part of the $\overline{\mathbf{126}}$ field. Even though
both of them have GUT scale masses we have
included them with the MSSM superpotential because their vevs lead to
light neutrino masses via the seesaw mechanism. 

The gauge coupling unification requires that the $\overline{\mathbf{126}}$
Higgs field acquires VEV at or close to the GUT scale.
We also need to introduce a $\mathbf{126}$ Higgs field to satisfy the
D-flat condition to maintain supersymmetry down to the weak scale.
Though the $\mathbf{126}$ Higgs field does not couple to the fermions,
the pair of Higgs doublets in the $\mathbf{126}$ mix with the doublets
arising from the 
other Higgs multiplets, since $\mathbf{126}$ couples to  the other Higgs
multiplets with non-zero coupling.  These five  pairs of Higgs doublets, 
$H_{u,d}^{10}$, $\Delta_{u,d}$, $\overline\Delta_{u,d}$, $A_{u,d}^s$ and
$A_{u,d}^{adj}$ 
are mixed and the  light pair of Higgs doublet can be  written as
\begin{eqnarray}
(H_u, \cdots) \!\!&=\!\!& (H_u^{10}, \Delta_u, \overline\Delta_u, A_u^s,
A_u^{adj}, \cdots)\ U_H, \\
(H_d, \cdots) \!\!&=\!\!& (H_d^{10}, \overline\Delta_d, \Delta_d, A_d^s,
A_d^{adj}, \cdots)\ V_H,
\end{eqnarray}
where $U_H$ and $V_H$ are unitary matrices. 
The superscripts $s$ and $adj$ stand for $SU(4)_c$ singlet and adjoint
pieces. We have temporarily ignored
the doublets that may arise from other multiplets in the theory such as
{\bf 210}. It is important to stress that in order to obtain one pair of
MSSM Higgs doublets from five pairs at GUT scale, one needs to do a fine
tuning of parameters. We have enough parameters in the Higgs
superpotential that this is possible to achieve. We have also checked that
we do not have any light color triplet fields.

 The results given below remain unchanged in their presence. The
Dirac mass matrices of quark and leptons are \begin{eqnarray}
M_u \!\!&=\!\!& M_{10} + M_{126} + M_{120}, \label{M_u} \\
M_d \!\!&=\!\!& r_1 M_{10} + r_2 M_{126} + r_3 M_{120}, \\
M_e \!\!&=\!\!& r_1 M_{10} - 3 r_2 M_{126} + A r_4 M_{120}, \\
M_\nu^D \!\!&=\!\!& M_{10} -3  M_{126} + A M_{120} \label{M_nu},
\end{eqnarray}
where the three mass matrices in the expression are given by
$M_{10}= h^* v_u (U_H)_{11}$, $M_{126}= c_1 f^* v_u (U_H)_{12}$, 
and $M_{120} = h'^* v_u ((U_H)_{14}+ c_2 (U_H)_{15})$
where $v_u$ is a vacuum expectation value of MSSM Higgs doublet $H_u$,
and $c_i$ are Clebsch-Gordon (CG)
coefficients.
The coefficients $r_i$ and $A$ are written as
\begin{eqnarray}
r_1 \!\!&=\!\!& (V_H)_{11}/(U_H)_{11} \cot\beta, \label{r1} \\
r_2 \!\!&=\!\!& (V_H)_{13}/(U_H)_{12} \cot\beta, \\
r_3 \!\!&=\!\!& ((V_H)_{14}+ c_2 (V_H)_{15})/((U_H)_{14}+ c_2
(U_H)_{15}) \cot\beta, \\
r_4 \!\!&=\!\!& ((V_H)_{14}-3 c_2 (V_H)_{15})/((U_H)_{14}-3 c_2
(U_H)_{15}) \cot\beta, \\
A \!\!&=\!\!& ((U_H)_{14}-3 c_2 (U_H)_{15})/((U_H)_{14}+ c_2 (U_H)_{15}),
\label{A}
\end{eqnarray}
where $\cot\beta$ is a ratio of vacuum expectation values of doublet
Higgs fields,
$\cot\beta = v_d/v_u$.
The Majorana mass matrices of left- and right-handed neutrinos prior to 
seesaw diagonalization are given by
\begin{equation}
M_L = f^* v_L , \quad M_R = f^* v_R,
\end{equation}
where $v_L$ and $v_R$ are vacuum expectation values of
$\overline\Delta_L$ and $\overline\Delta_R$,
respectively. 
As already mentioned, since $v_R$ is expected to be close to the GUT
scale, this implies that $v_L$ is 
$\sim$$v_{\rm weak}^2/(\eta M_{GUT})$ $\ll v_{\rm weak}$,
where $\eta$ is a coupling constant in the  Higgs potential.
The Majorana mass matrix of the heavy right handed neutrino is
proportional to $M_{126}$.
The light neutrino mass matrix is given by the mixed type II seesaw
formula,
\begin{equation}
{\cal M}_\nu^{\rm light} = M_L - M_\nu^D M_R^{-1} (M_\nu^D)^T.
\end{equation}
As discussed in earlier papers\cite{goran,goh}, there are regions of the
parameter space in the theory where the first term will dominate; we will
call this the pure type II seesaw case.
If on the other hand, we consider the parameter space where the second
term is dominant we will call this type I seesaw. The bulk of our results
will be for the pure type II case.

\section{Parity invariance and a predictive model for neutrinos} 
In order to see if the model is predictive for neutrinos, let us count
the number of parameters in the theory.
In the basis, where $M_{126}$ is real and diagonal,
there are 3 real parameters in $M_{126}$, 
6 complex parameters in $M_{10}$ and 3 complex parameters in $M_{120}$.
We also have 5 complex parameters in the Eqs.(\ref{r1}-\ref{A}) as well as
the vevs of the 
Altogether, there are 31 real parameters in the fermion sector,
and, therefore, we do not have any prediction for the neutrino mixings.

In order to be  predictive in  the leptonic sector of the model without
imposing any flavor symmetry, we require the theory to be invariant under 
 a parity symmetry. As we will see, it makes the Dirac mass matrices
Eqs.(\ref{M_u}-\ref{M_nu}) hermitian and leaves a total of
 17 real parameters in the fermion sector making the model predictive. If
we further require that the {\bf 120} Higgs field has a mass much
higher than the GUT scale, its only manifestation is as an effective
dimension four term in the superpotential. The reduces the number of
parameters to 15 increasing the predictive power of the model. We explore
both the cases with 17 and 15 parameters in a subsequent section.

We now define the parity transformation in the $G_{422}$ basis.
We write SU(4) indices by $\mu$, $\nu$, SU(2)$_L$ indices by $\alpha$,
$\beta$
and SU(2)$_R$ indices by $\dot\alpha$, $\dot\beta$.
The SO(10) spinor $\psi$ and $\chi$ are decomposed as
\begin{equation}
\psi = \psi_{\mu\alpha} + \psi^\mu_{\dot\alpha}, \qquad
\chi = \chi_{\mu\alpha} + \chi^\mu_{\dot\alpha}.
\end{equation}
Bi-doublet Higgs fields in the $\mathbf{10}$, $\overline\mathbf{126}$ and
$\mathbf{120}$
are written as 
$H_{\alpha\dot\alpha}$, $\overline\Delta_\mu{}^\nu{}_{\alpha\dot\alpha}$,
$A_{\alpha\dot\alpha}$, and $A_\mu{}^\nu{}_{\alpha\dot\alpha}$.
Then the Yukawa interactions are written in the following (up to overall
factors)
\begin{eqnarray}
h\ \psi \chi H \!\!&=\!\!& h (\psi_{\mu\alpha} \chi^\mu_{\dot\alpha}
                                 + \psi^\mu_{\dot\alpha}
\chi_{\mu\alpha}) H^{\alpha\dot\alpha}+ \cdots
\\
f\ \psi \chi \overline\Delta \!\!&=\!\!& f(\psi_{\mu\alpha}
\chi^\nu_{\dot\alpha}
                                + \psi^\nu_{\dot\alpha} \chi_{\mu\alpha})  
\overline\Delta_\nu{}^\mu{}^{\alpha\dot\alpha}+ \cdots
\\
h'\ \psi \chi A \!\!&=\!\!& h'(\psi_{\mu\alpha} \chi^\mu_{\dot\alpha}
                                - \psi^\mu_{\dot\alpha} \chi_{\mu\alpha}) 
                                 A^{\alpha\dot\alpha}+ 
                                 c_2 h'(\psi_{\mu\alpha} \chi^\nu_{\dot\alpha}
                                - \psi^\nu_{\dot\alpha} \chi_{\mu\alpha}) 
                                 A_\nu{}^\mu{}^{\alpha\dot\alpha}+ \cdots
\end{eqnarray}
where $c_2$ is a CG coefficient. 
The Lagrangian is written as
\begin{equation}
{\cal L} = \int d^2 \theta\ W + \int d^2\bar\theta\ \overline W
\end{equation}
and 
\begin{equation}
{\cal L} = \int d^2 \theta \ h (\psi_{\mu\alpha} \chi^\mu_{\dot\alpha}
                 + \psi^\mu_{\dot\alpha} \chi_{\mu\alpha}) H^{\alpha\dot\alpha}
          + \int d^2 \bar\theta \ h^* ((\psi_{\mu\alpha})^* 
(\chi^\mu_{\dot\alpha})^*
                  + (\psi^\mu_{\dot\alpha})^*
(\chi_{\mu\alpha})^*) (H^{\alpha\dot\alpha})^* + \cdots
\end{equation}
We consider the symmetry under the following parity transformation,
\begin{eqnarray}
\!\!&\!\!&\!\! \psi_{\mu\alpha} \leftrightarrow (\psi^{\mu {\dot\alpha}})^*, 
\quad d^2\theta \leftrightarrow d^2\bar\theta.
\end{eqnarray}
Of course, $\chi$ is also transformed in same manner.

In the Higgs sector, the transformations of the $(\mathbf{1,2,2})$ and 
$(\mathbf{15,2,2})$ sub-multiplets under $G_{422}$ are:
 \begin{eqnarray}
\!\!&\!\!&\!\!          H^{\alpha\dot\alpha} \leftrightarrow
(H_{\alpha\dot\alpha})^*, \quad                 
 \overline\Delta_\nu{}^\mu{}^{\alpha\dot\alpha} \leftrightarrow 
(\overline\Delta_\mu{}^\nu{}_{\alpha\dot\alpha})^*,
\quad
A^{\alpha\dot\alpha} \leftrightarrow 
(A_{\alpha\dot\alpha})^*   ,
\quad
A_\nu{}^\mu{}^{\alpha\dot\alpha} \leftrightarrow 
(A_\mu{}^\nu{}_{\alpha\dot\alpha})^* 
 \label{parity-trans}.
\end{eqnarray}
A consequence of the parity symmetry (\ref{parity-trans}),
is that the coupling matrices $h$ and $f$ real and symmetric and $h'$
antisymmetric and 
imaginary; the parameters $r_i (i=1,2,3,4)$ and $A$ in the
Eqs.(\ref{r1}-\ref{A}) are real. This considerably reduces the number of
parameters in the theory and further makes the mass matrices for all
charged fermions hermitian.

Let us clarify our motivation for introducing the $\mathbf{120}$ Higgs field.
When the  $\mathbf{120}$ Higgs field is absent,
the fermion mass matrices are complex symmetric matrices in the absence 
of the parity symmetry and
we have the following relation in the pure type II case 
\begin{eqnarray}
&&{\cal M}_\nu \propto M_d - r_1 M_u = 
U(V D_d V^T - r_1 D_u)U^T
\nonumber \\
&& 
\simeq U \left( 
\begin{array}{ccc}
m_d e^{i \phi_d} + V_{us}^2 m_s e^{i \phi_s} & V_{us} m_s e^{i \phi_s}
& V_{ub} m_b \\
V_{us} m_s e^{i \phi_s} & m_s e^{i \phi_s} & V_{cb} m_b \\
V_{ub} m_b & V_{cb} m_b & m_b - r_1 m_t  
\end{array}
\right) U^T,
\label{relation0}
\end{eqnarray}
where $m_c$ and $m_u$ contributions in (2,2) and (1,1) elements 
are omitted, and $\phi_d$ and
$\phi_s$ are complex phases in the diagonal matrix, $D_d$. 
If $M_e$ is close to a diagonal matrix in the basis where $M_u$ is
diagonal,
the maximal atmospheric mixing can be easily obtained when 
the (3,3) element is suppressed such that
$|m_s e^{i \phi_s} -(m_b - r_1 m_t)|\ll 2 V_{cb} m_b$.
This suppression of (3,3) element 
is related with other observed facts
such as
bottom and tau mass convergence
at GUT scale and 
$\Delta m^2_{sol}/\Delta m^2_A \gg O(m_s^2/m_b^2)$.
Assuming that the atmospheric mixing is maximal,
we obtain the neutrino mass matrix, Eq.(\ref{relation0}), as
\begin{equation}
\simeq U U_{23} \left( 
\begin{array}{ccc}
m_d e^{i \phi_d} + V_{us}^2 m_s e^{i \phi_s} & 
-(V_{ub} - V_{us} V_{cb}) m_b/\sqrt2 
& (V_{ub}-V_{us} V_{cb}) m_b/\sqrt2 \\
-(V_{ub}-V_{us}V_{cb}) m_b/\sqrt2 & \epsilon m_0 & 0 \\
(V_{ub}-V_{us}V_{cb} ) m_b/\sqrt2 & 0 & m_0  
\end{array}
\right) U_{23}^T U^T,
\end{equation}
where $m_0$ and $\epsilon m_0$ are eigenvalues of (2-3) block,
and $\epsilon^2 \sim \Delta m^2_{sol}/\Delta m^2_A$ and $m_0 \sim 2 m_s$.
Thus, the solar mixing and 13 mixing are proportional to 
$|V_{ub}/V_{cb}-V_{us}|$
and $\tan 2\theta_{sol} \sim 
\tan 2\theta_{13}/\epsilon$.
Therefore, those mixing angles also depend on the KM phase, 
$\delta_{\rm KM}$.
Since $V_{ub} = |V_{ub}| e^{-i \delta_{\rm KM}}$,
the KM phase in the first quadrant gives a smaller value for solar mixing angle
rather than  in the second quadrant.
In order to obtain the proper value of solar mixing angle, we have to
choose a smaller value of $\epsilon$.
However, in the model without {\bf 120}, the $\epsilon$ parameter,
which is a function of the strange quark mass, is
constrained due to the fitting of three charged-lepton masses 
(especially electron mass),
and we do not have proper  fitting of the solar mixing angle data
in the case where the KM phase is in the first quadrant.
We can verify the situation in a precise analysis in the pure type II case
\cite{goh1}.
In the type I case, things are more complicated, but it has been shown that
it is not possible to fit  the  neutrino oscillation data in this model
with the above
 minimal Higgs choice \cite{mimura}. The mass squared ratio is constrained
due to the charged-lepton mass fitting and
it cannot be small enough when the KM phase is in the
first quadrant. The mass squared ratio is a free parameter in the model 
with {\bf 120} since the additional parameter $A$ 
in the sum rules Eqs.(\ref{M_u}-\ref{M_nu}) can fit the electron mass,
 and therefore we can explain a smaller KM phase. 
Thus, we  employ the $\mathbf{120}$ Higgs field to explain
the large solar mixing angle along with the KM phase in the first quadrant.
Interestingly, even though we have introduced a new Higgs field,
the number of parameters is less than the minimal Higgs choice with most 
general CP phases due to the constraint of parity symmetry.

We further note that if the {\bf 120} field is heavier than the GUT scale,
its effect on the physics at the GUT scale comes from a higher dimensional
operator of the form $\psi\Gamma\Gamma\Gamma\psi H \Phi/M$, where
$\Phi$ is a {\bf 210} Higgs field, so that
we get the relation $r_1=r_3=r_4$.  In other words, the choice $r_1=r_3=r_4$ is
not an adhoc choice but can be guaranteed in a natural manner.

We now note that in the presence of the parity symmetry, since all the
mass
matrices are hermitian and also the $\mu$-term and the gluino masses are
real, the most dangerous graphs contributing to large electric dipole
moment of the neutron are absent\cite{rasin}. Therefore this model has the
potential to solve the so-called SUSY CP problem.

We also wish to recognize that the $Z_2$ CP symmetry we impose is 
broken at the GUT scale by the VEVs of $\mathbf{126}$, 
$\mathbf{210}$ and $\mathbf{45}$ Higgs fields which break SO(10) symmetry.
The light MSSM doublet Higgs fields are no more CP eigenstates,
and thus there is no cosmological domain wall problem at weak scale in the model.

\section{Near Bi-Maximal Solution for Neutrino Oscillation}
As already noted in section 3, under the parity symmetry,
$M_{10}$ and $M_{126}$ are real symmetric matrices,
$M_{120}$ is a pure imaginary anti-symmetric matrix,
and the coefficients $r_i$ and $A$ are real parameters
in the Eqs.(\ref{M_u}-\ref{M_nu}). We can therefore rewrite the
 charged-lepton and Dirac neutrino mass matrices in terms of the other
mass matrices as follows:
\begin{eqnarray}
M_e &\!\!=&\!\! c_u \ {\rm Re} M_u + c_d \ {\rm Re} M_d + i Ar_4 \ {\rm
Im} M_u  , \\
M_\nu^D &\!\!=&\!\! {\rm Re} M_u + \frac{3+c_d}{r_1} \ {\rm Re} (M_d -
r_1 M_u) + i A\ {\rm Im} M_u ,
\end{eqnarray}
where $c_u/(1-c_d) = r_1$, $-c_u/(3+c_d)=r_2$.
The up- and down-type quark mass matrices are hermitian, and
are written as
\begin{equation}
M_u = U  D_u U^\dagger, \quad
M_d = U V  D_d V^\dagger U^\dagger,
\end{equation}
where
$D_u={\rm diag}(\pm m_u, \pm m_c, m_t)$, $D_d={\rm diag}(\pm m_d, \pm
m_s, m_b)$
and $V$ is the CKM matrix, and $U$ is a unitary matrix.
We note that $m_t$ and $m_b$ component in the $D_u$ and $D_d$ can be
made to be positive
without loss of generality.
Because of the parity symmetry,
$M_d-r_3 M_u$ must be a real symmetric matrix.
We fix the flavor basis as
\begin{equation}
M_d - r_3 M_u = U (V D_d V^\dagger - r_3 D_u) U^\dagger = {\rm diag.}
\end{equation}
The unitary matrix $U$ is determined by $r_3$, up to phase matrix $P$,
\begin{equation}
U = P \bar U, \quad P = {\rm diag} (e^{i \phi_1}, e^{i \phi_2},1)
\end{equation}

The parameters are now 6 quark masses (with signatures), 3 mixing and 1
KM phase in the CKM matrix,
and the coefficients $c_d$, $c_u$, $\phi_1$, $\phi_2$, $r_3$, $r_4$, $A$.
There are 17 parameters in all.
For example, the three charged-lepton masses can be used to determine
 $c_d$, $c_u$, $A r_4$.
The remaining 4 parameters give the neutrino oscillation parameters.

In the case where the {\bf 120} Higgs field is heavier than the GUT scale,
it manifests itself as an effective dimension four operator of the
form $\psi\psi H \Phi/M$. As a result, the Higgs doublets in {\bf
120} are decoupled. This leads to a reduction in the number of mixing 
parameters. This translates into the
the relation $r_1=r_3=r_4$ since the vev ratio that defines
$r_{1,3,4}$ is the same. In this case, there remain only two
parameters describing the neutrino sector. They can be determined by two
of the parameters from the
neutrino oscillation data and the remaining neutrino parameters can then
be predicted.

Interestingly, if we have the relation $r_1=r_3$,
the matrix $M_d - r_3 M_u$ is proportional to the light neutrino mass
matrix $M_L$
and the diagonalizing matrix $U$ become close to MNSP matrix in the pure
type II case.
Therefore, if $r_1=r_3 \sim m_b/m_t$ and the (3,3) element of $M_d-r_1
M_u$ is suppressed,
we have a large atmospheric mixing.
The mass squared ratio is of the order of $10^{-2}$, which is the right
order seen in the experiment,
only if $r_1 \sim m_b/m_t$ (otherwise, the mass squared ratio become
$O(m_c^2/m_t^2)$ or $O(m_s^2/m_b^2)$).
Furthermore, since the (3,3) element of $M_{126}$ is suppressed
for that choice of $r_1$, the bottom-tau mass unification is satisfied
and it is consistent with the renormalization group flow for the case of
$\tan\beta \sim 50$.

Now let us study the prediction of the model in the case where
$r_1=r_3=r_4$.
In this case, we have 15 parameters in the model.
After fixing the quark masses (with signatures) and the CKM parameters,
we are left with 5 parameters, $c_d$, $A$, $\phi_1$, $\phi_2$, and $r_1$.
Since mass squared ratio is a function of $r_1$,
we can fix the parameter $r_1$ by the experimental value of $\Delta
m^2_{sol}/\Delta m^2_A$.
The three charged-lepton masses can be used to fix  $c_d$, $A$, and
$\phi_2$. As a result,
the neutrino oscillation parameters, $\theta_A$, $\theta_{sol}$,
$|U_{e3}|$,
and one CP phase $\delta_{MNSP}$ are predicted by only one phase
parameter $\phi_1$.
Interestingly, the atmospheric mixing $\theta_A$ doesn't depend on the
phase $\phi_1$ very much,
and the $\theta_A$ is really predicted when we fix the quark masses
and mass squared ratio of light neutrino. 

It should also be noted that the other arbitrariness in the model is due
to the choice of the signs of different fermion masses, since the sign of
a fermion mass is unobservable. We find that only for the two choices of
the signs given below, we obtain acceptable solutions: 
\begin{eqnarray}
({\rm a}) && D_u = {\rm diag}(\pm,-,+), \quad D_d = {\rm diag} (-,+,+),
\quad D_e = {\rm diag} (\pm,-,+),
\label{signature:a} \\
({\rm b}) && D_u = {\rm diag}(\pm,-,+), \quad D_d = {\rm diag} (+,+,+),
\quad D_e = {\rm diag} (\pm,+,+).
\label{signature:b}
\end{eqnarray}
The solutions we present correspond only for these two choices of
signatures.

In Fig.1, we show the prediction of the atmospheric mixing $\sin^2
2\theta_A$
as a function of the mass squared ratio.
The lines (a) and (b) in the figure are for the set of quark and lepton
mass signatures given above
which give acceptable solutions for neutrinos.
We can obtain large atmospheric mixing angles with the proper
choice of quark masses in their allowed range, mixings and KM phase
\cite{PDG} with the choice of the set of signatures, especially for case
(a).
The most important input parameter for obtaining a large
atmospheric mixing is the strange quark mass.
We show the strange mass dependence of $\theta_A$ in Fig.2 in the case
where the mass squared ratio $R\equiv \Delta m^2_{\odot}/\Delta m^2_A$ is
0.03.
The strange mass in the figure is the running mass at 1 GeV.
In order to obtain the experimental constraint $\sin^2 2\theta_A > 0.9$,
we need the parameter region where the strange mass has a larger value.
The bottom quark mass dependence is not negligible,  and a
larger value of bottom mass
is preferred to obtain maximal atmospheric mixing.

After fixing all quark and lepton data and also the mass squared ratio, $R$,
we can fit the solar mixing angle by choosing the free phase parameter
$\phi_1$.
Then, $|U_{e3}|$ and $\delta_{MNSP}$ are predicted.
In Fig.3, we show the correlation between solar mixing
$\tan^2 \theta_{sol}$
and $|U_{e3}|$ by varying the phase parameter $\phi_1$.
We give two lines for the cases (a) and (b),
and the two lines correspond to  different mass squared ratios, $R
=0.02$ and $R= 0.07$.
From the figure, we can see that $\Delta m^2_{\odot}/\Delta m^2_A
=0.07$ is not
favored in the $3\sigma$ range of the experimental data of solar
neutrino and $U_{e3}$ for the case (a).
In Fig.4, we present the prediction of the $\sin \delta_{MNSP}$
in the case where $R=\Delta m^2_{\odot}/\Delta m^2_A =0.02$.
The CP phase can be of any value in the range of the experimental data
of solar neutrino.
If we restrict the mass signatures to the (a) case,
$\sin \delta_{MNSP}$ could be predicted to be $\pm 0.9$.

The most interesting feature of this model is
the prediction of $|U_{e3}|$.
In Fig.5, we show the bottom quark mass dependence of the $|U_{e3}|$
prediction.
The bottom mass is defined as a running mass at $m_b$.
Since a large SUSY correction to the bottom quark mass can be induced in
the large $\tan\beta$ case,
the running bottom mass can be large,
and the larger bottom mass gives smaller value of $|U_{e3}|$.
In a similar way, larger $\tan\beta$ predicts smaller $|U_{e3}|$ 
since the larger $\tan\beta$ gives larger bottom quark mass at GUT scale.
In Fig.6, we show the plot of $|U_{e3}|$.
Each point is dotted for different quark mass and mixings
which are randomly generated in the experimentally allowed region.
We can see that the model predicts a lower limit for $|U_{e3}|$ of
about 0.1.

In Fig.7, we can see the KM phase dependence of the prediction of the
model.
The lines in the figure are drawn by changing $\phi_1$ for different values of  KM
phases.
The mass squared ration is set to be 0.02.
For a smaller KM phase, the lines shift to the smaller solar mixing.
In this model,
the experimentally allowed solar mixing can be obtained
in the first quadrant KM phase, contrary to the minimal model without
the $\mathbf{120}$ Higgs field.
We can see that a smaller value of $|U_{e3}|$ can be allowed for the
larger values of KM phase.

Since all the parameters of the model are now determined, it can be used 
to make other predictions. As an example, we have calculated the
magnitude of lepton flavor violating process $\mu\rightarrow e +\gamma$ in
the model (see Fig.\ref{Fig.8}). Note that the predictions are in the
range
currently being probed by experiments\cite{meg}. We use the mSUGRA universal boundary 
conditions at the GUT scale i.e. $m_0$ (universal scalar mass),
$m_{1/2}$ (universal gaugino mass),$A_0$ (universal trilinear mass ). The other two 
parameters are 
 sign of $\mu$ and $\tan\beta$. 
The dots in the  plots are  produced for various model points generated by fitting 
the fermion masses and mixing
angles. We can see that the BR of $\mu\rightarrow e\gamma$ can be large for smaller 
values of $m_{1/2}$. 
The lightest neutralino is the dark matter candidate in this model
and we satisfy the 2$\sigma$ range of the recent relic density constraint 
$\Omega_{\rm CDM}=0.1126^{+0.008}_{-0.009}
$\cite{wmap} in the parameter space. When we satisfy the relic density
constraint, the $m_0$ gets determined. We choose $A_0=0$ and $\mu>0$. The right
handed masses have hierarchies and therefore get decoupled at different
scales. The flavor-violating pieces present
in $Y_{\nu}$ and $f$ induces flavor violations into the charged lepton couplings and 
into
the soft SUSY breaking masses e.g.  $m^2$ terms etc. Also, an additional 
 symmetry  between the GUT and the $v_R$ scale (type I) helps to induce flavor 
violation\cite{mimura}. The electric dipole moment of
electron is smaller than the experimental reach in the range of parameter space showed 
in
the figure.

We comment that in the case of type I seesaw, 
the large mixing solution is a sharp resonance solution
and the solution is not stable to predict the mixings
contrary to the type II seesaw.

\section{Discussion}
In this paper we consider the prediction of neutrino masses and mixings
for an SO(10) model where fermion masses
receive contribution from the presence of three Higgs multiplets {\bf 10},
{\bf 120} and $\overline{\bf 126}$. We impose a parity symmetry on the
model, so that it has very few parameters which enables prediction of two
mixing
angles and all the CP phases in the neutrino mixing matrix. The advantage
of this model over the most minimal SO(10) model is that now the CKM phase
is in the first quadrant as required by the standard model analysis of all
observed hadronic CP violation. We also wish to emphasize that this one of
the few models in the literature that can predict leptonic CP phases.

As far as experimental tests of this
model are concerned, the parameter
$U_{e3}\equiv \sin\theta_{13}$ is predicted to be large like the most minimal
SO(10) model\cite{goran,goh} but is somewhat smaller i.e. $U_{e3}\geq
0.1$. This can be tested in the next round of planned long baseline
experiments. We also predict that
the Dirac phase for neutrinos can be maximal. Furthermore the model has
the potential to solve 
SUSY CP problem due to the fact that all fermion masses are
hermitian. The model also predicts observable amount of LFV in muon decay.

\section*{Acknowledgments}
This work of B.D. and Y.M. is supported by 
the Natural Sciences and Engineering Research Council of Canada and 
the work of R. N. M. is supported by the National Science Foundation
Grant No. PHY-0354401.

Note added: After this work was completed and was being prepared for
publication, two papers appeared which also
include the effect of {\bf 120} Higgs field on fermion
masses\cite{frigerio,ng} in the SO(10) model.


\begin{figure}[p]
\centering
\includegraphics*[angle=0,width=11cm]{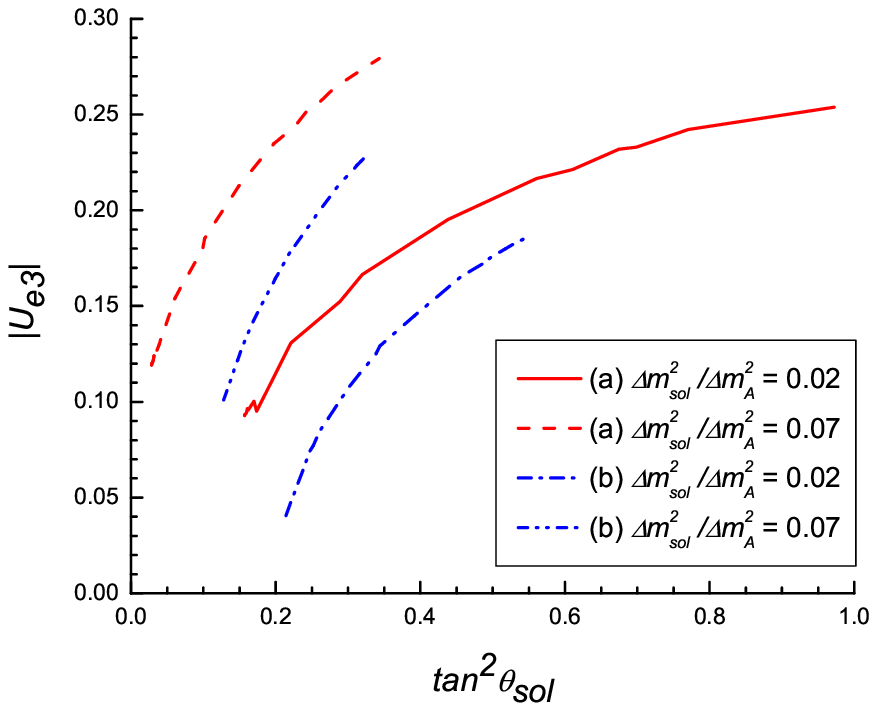}
 \caption{The atmospheric mixing angle is plotted as a function of mass
squared ratio.
 Predictions for different set of mass signatures (a) and (b) in
Eqs.(\ref{signature:a}-\ref{signature:b})
are given.}
\label{Fig.1}
\end{figure}
\begin{figure}[p]
\centering
\includegraphics*[angle=0,width=11cm]{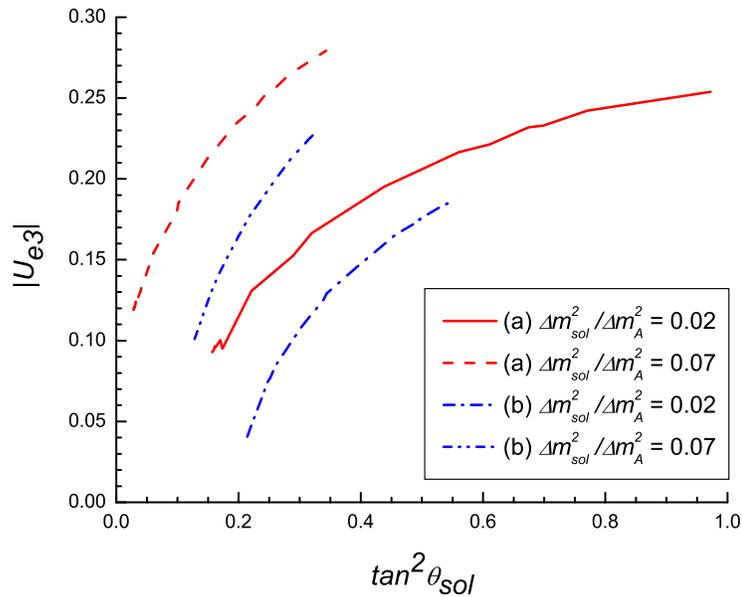}
 \caption{The atmospheric mixing angle is plotted as a function of strange
quark mass.
 The strange quark mass is given as a running mass at 1 GeV. The mass
squared ratio
is 0.03 in each set of mass signature, (a) and (b).}
\label{Fig.2}
\end{figure}

\begin{figure}[p]
\centering
\includegraphics*[angle=0,width=11cm]{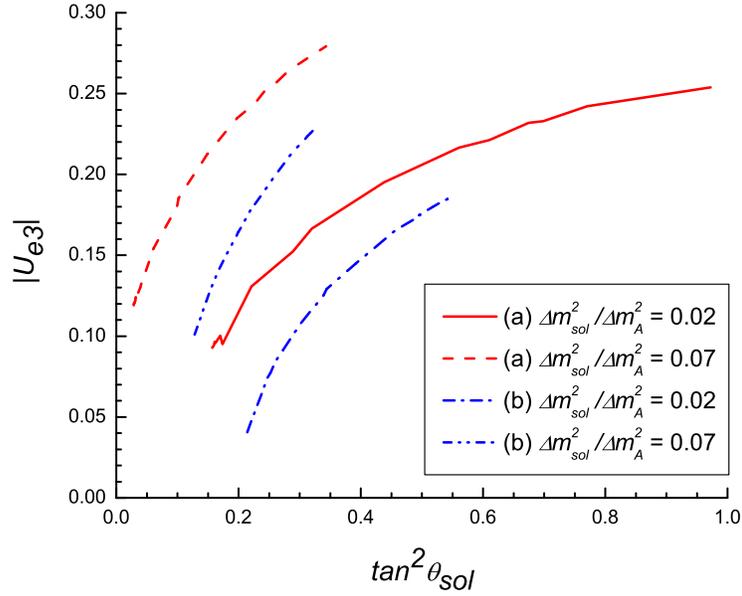}
\caption{The relation between solar mixing angle and $|U_{e3}|$ is plotted.
Each line is plotted by varing a free phase parameter $\phi_1$.
The experimentally allowed region in $3\sigma$ of recent data fitting
is $0.3 < \tan^2\theta_{sol} <0.6$ and $|U_{e3}|<0.26$ \cite{fogli}.}
\label{Fig.3}
\end{figure}
\begin{figure}[p]
\centering
\includegraphics*[angle=0,width=11cm]{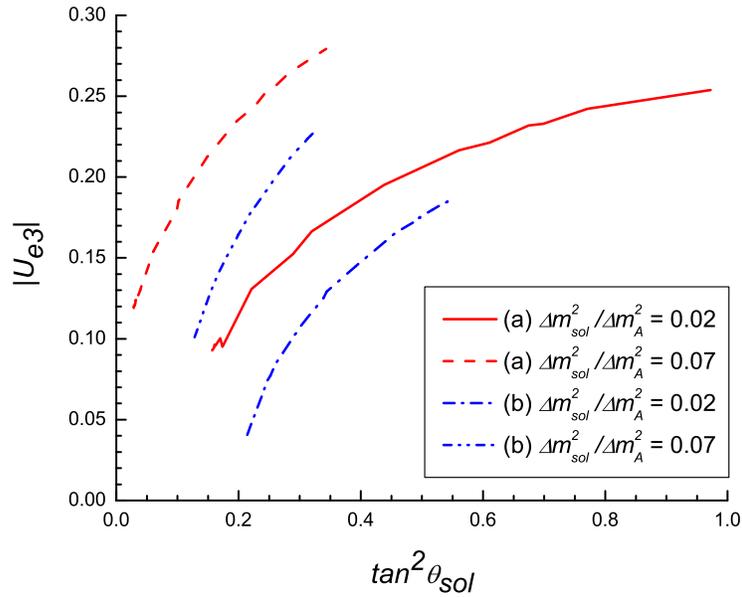}
 \caption{The prediction of MNSP phase is plotted as a function of the
solar mixing angle.
 These lines (a) and (b) are plotted in the case $\Delta m^2_{sol}/\Delta
m^2_A = 0.02$
for different mass signature.}
\label{Fig.4}
\end{figure}
\begin{figure}[p]
\centering
\includegraphics*[angle=0,width=11cm]{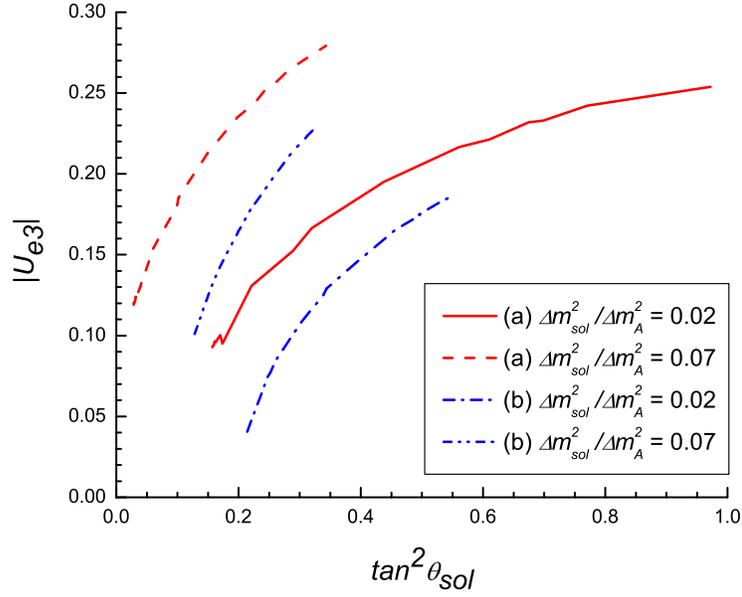}
 \caption{The relation between solar mixing angle and $|U_{e3}|$ is
plotted as dots
 for different bottom quark masses. The bottom quark mass is given as a
running mass at $m_b$.
This plot is given in the case of mass signature (a) 
and mass squared ratio is 0.02.}
\label{Fig.5}
\end{figure}
\begin{figure}[p]
\centering
\includegraphics*[angle=0,width=11cm]{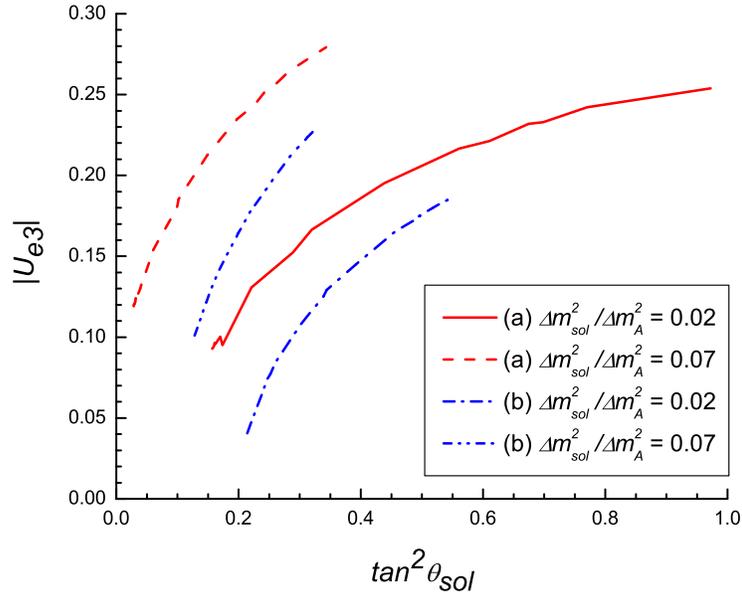}
 \caption{The prediction of $|U_{e3}|$ and solar mixing angle is plotted
as dots for randomly
 generated quark masses (with signature) and mixings in the experimantally
allowed region.
The lower bound of $|U_{e3}|$ exists in this model.}
\label{Fig.6}
\end{figure}
\begin{figure}[p]
\centering
\includegraphics*[angle=0,width=11cm]{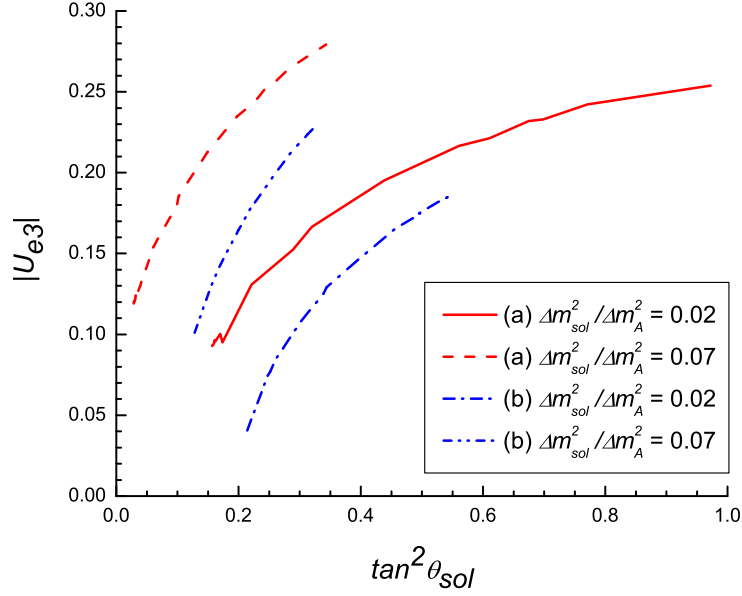}
\caption{The relation between solar mixing angle and $|U_{e3}|$ is plotted for
various KM phases.
Each plot is given in the case of mass signature (a)
and mass squared ratio is 0.02.}
\label{Fig.7}
\end{figure}
%
%

\begin{figure}[tbp]
    \centering
    \includegraphics*[angle=0,width=10cm]{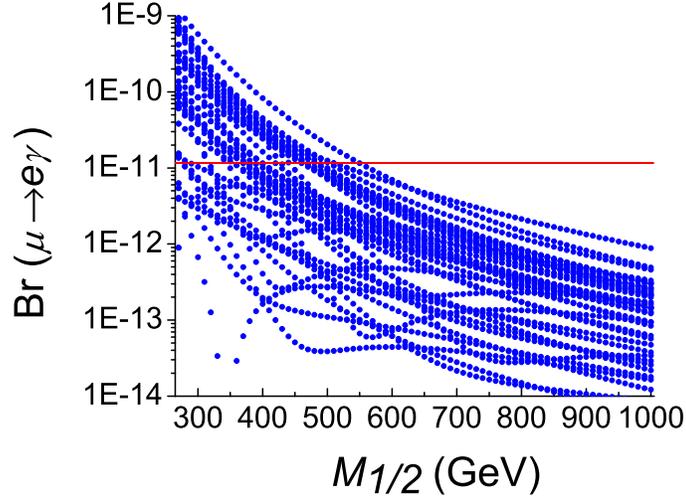}
    \vspace{0.1cm}
    \caption{ The BR[$\mu\rightarrow
e\gamma$] is plotted as a function of $m_{1/2}$ for  $\tan\beta=10$ in  type II.}
\label{Fig.8}
\end{figure}
\end{document}